\documentclass[12pt,a4]{article}

\usepackage{graphicx}
\usepackage{color}

\newcommand{\rf}[1]{(\ref{#1})}
\newcommand{\beq}{\begin{equation}}
\newcommand{\eeq}{\end{equation}}
\newcommand{\bea}{\begin{eqnarray}}
\newcommand{\eea}{\end{eqnarray}}

\newcommand{\e}{{\rm e}}

\newcommand{\equ}{\!=\!}
\newcommand{\plu}{\!+\!}
\newcommand{\mi}{\! - \!}

\newcommand{\Lam}{\Lambda}

\newcommand{\Om}{\Omega}

\newcommand{\kp}{\kappa}

\newcommand{\oq}{\frac{1}{4}}

\def\void{}
\def\labelmark{}

\newenvironment{formula}[1]{\def\labelname{#1}
\ifx\void\labelname\def\junk{\begin{displaymath}}
\else\def\junk{\begin{equation}\label{\labelname}}\fi\junk}%
{\ifx\void\labelname\def\junk{\end{displaymath}}
\else\def\junk{\end{equation}}\fi\junk\labelmark\def\labelname{}}

{\ifx\void\labelname\def\junk{\end{array}\end{displaymath}}
\else\def\junk{\end{array}\right.\end{equation}}
\fi\junk\labelmark\def\labelname{}\def\junk{}
}

\newcommand{\beqv}{\begin{formula}{}}

\vspace{12pt}

\begin{document}

\rightline{\today}

\begin{center}
\vspace{24pt}
{ \Large \bf Easing the Hubble constant tension}

\vspace{24pt}

{\sl J.\ Ambj\o rn}$\,^{a,b}$ and {\sl Y.\ Watabiki}$\,^{c}$

\vspace{10pt}

{\small

$^a$~The Niels Bohr Institute, Copenhagen University\\
Blegdamsvej 17, DK-2100 Copenhagen \O , Denmark.\\
email: ambjorn@nbi.dk
\vspace{10pt}

$^b$~Institute for Mathematics, Astrophysics and Particle Physics
(IMAPP)\\ Radbaud University Nijmegen, Heyendaalseweg 135, 6525 AJ, \\
Nijmegen, The Netherlands

\vspace{10pt}

$^c$~Tokyo Institute of Technology,\\ 
Dept. of Physics, High Energy Theory Group,\\ 
2-12-1 Oh-okayama, Meguro-ku, Tokyo 152-8551, Japan\\
{email: watabiki@th.phys.titech.ac.jp}

}

\end{center}

\vspace{24pt}

\begin{center}
{\bf Abstract}
\end{center}

\noindent
We show how a modified Friedmann equation, originating 
from   a model of the universe built from a certain $W_3$ algebra, has the potential to explain
the difference between the Hubble constants extracted from CMB data and from supernova data.

\newpage

\section{Introduction}\label{intro}

The model of our Universe introduced in \cite{aw1} is an attempt to explain 
how time and space emerged in our present universe, starting from a string field theory based on
a certain $W_3$ algebra. It led in addition (under certain natural assumptions) to a 
modified Friedmann equation \cite{aw2}.
An appealing feature of this Friedmann equation is that it needs no cosmological term, but that it nevertheless results in 
an exponentially expanding universe at late times and that this exponential expansion is linked to the creation of baby universes 
and wormholes in the  quantum universe. In this way the quantum aspects of gravity at the smallest distances become
linked to the largest distances of our classical universe. We refer to \cite{aw1} for a detailed discussion of these aspects, but let us 
for the convenience of the reader just outline the general picture emerging from \cite{aw1}: by a breaking of the mentioned
$W_3$ symmetry,  universes can be created. They have both cosmological constants and a new constant related to the 
creation of baby universes. It is dominantly two-dimensional universes with different ``flavors'' (coming from the $W_3$ algebra)
which expand and they might expand very fast (faster than exponentially) to macroscopic scales. During this expansion
the interaction between universes with different flavors turns these two-dimensional universes into a higher dimensional universe,
the dimension of which is determined by the symmetry breaking of the $W_3$ algebra. Possible dimensions of the resulting 
spacetime are 3, 4, 6 and 10. The expansion triggered by the cosmological constant is then stopped because the Coleman
mechanism is operating,  and we imagine that it will also be operating  at all later times. The appearance of dimensions 3, 4, 6 and  10 is of course tantalizing and suggests the possibility that our model could be equivalent to a superstring theory where all matter fields are integrated out.

The purpose of this article is to show that without further assumptions than already present in \cite{aw2}  our modified 
Friedmann equation can explain why  one obtains different values for the Hubble constant $H_0$ when  using 
CMB data \cite{planck} and supernova data \cite{SNIa}. Before doing that, in order to avoid misunderstanding, let us clarify
which cosmological problems we {\it do not address}, namely the hierarchy problem and the naturalness problem, and what problem
we {\it do address}, namely the $H_0$ problem.

The hierarchy problem is that the observed value of dark energy is much 
too small compared to the Planck scale which will appear in any 
calculation involving quantum field theory interacting with gravity. We will assume that this problem has been solved 
by the Coleman mechanism and that the  value of dark energy we observe today comes entirely from our new parameter $B$.
This parameter is not influenced by the Coleman mechanism, but we will not here try to explain its actual value. In principle it might be possible since it is related to the creation of baby universes in our ``real'' theory based on $W_3$ algebra, which underlies 
our modified Friedmann equation. However, presently we are not able to follow the renormalization flow of this coupling constant 
from the creation of our universe to present time. Thus we cannot explain the ``smallness'' of our parameter $B$, just like 
the smallness of $\Lam$ cannot be explained from first principles in the standard $\Lam$CMD cosmology.

The naturalness problem is that after symmetry breaking of any kind, 
whether it is at electroweak scale or QCD scale or any similar scale, 
the  contribution to dark energy will be very large compared to the present small value and some unknown fine tuning has to take place to accommodate the present observed value of dark energy. We do not try to solve this problem, but again simply assume that 
the Coleman mechanism is operating and the cosmological constant coming from such contributions can be set to zero. 
Since our model provides us with a new mechanism for the expansion of the universe at late time, 
we do not need a non-zero cosmological constant.

A third cosmological problem is the $H_0$ problem: if one believes in the direct, local  observation of $H_0$, which is obtained 
by the cosmic ladder of Cepheid-SNIa standard candles, it differs by $5 \, \sigma$ from the estimated $H_0$ value obtained
from the CMB data, obtained by using the $\Lam$CDM model to extrapolate from the time of last scattering to present days time.
This is the problem to which our model provides a solution, since our modified Friedmann equation leads to a different extrapolation
from the time of last scattering to present days time than provided by the $\Lam$CDM model.

 \section{The modified Friedmann equation}
 
 The (first) Friedmann equation is
 \beq\label{2.1}
 \left(\frac{\dot{a}}{a}\right)^2 = \frac{\kp \rho + \Lam}{3},\qquad \dot{a} := \frac{da(t)}{dt},
 \eeq
 where $\kp = 8\pi G$ and where $G$ and $\Lam$ denote the gravitational and the cosmological constants. $a(t)$ denotes
 the scale factor in a universe assumed homogeneous and isotropic with matter density $\rho$ and we have assumed
 the spatial curvature  is zero, i.e.\ $\Om_{\rm K} =0$. In \cite{aw2} we showed that our modified Friedmann equation is 
 \beq\label{2.2}
  \left(\frac{\dot{a}}{a}\right)^2 = \frac{\kp \rho }{3} + \frac{B a}{\dot{a}} \frac{(1+3F(x))}{F^2(x)},\quad x := \frac{B a^3}{\dot{a}^3},
  \quad F^2(x)-F^3(x) = x. 
  \eeq
 In eq.\ \rf{2.2} we have assumed that $\Lam =0$, in accordance with our above mentioned assumption that the Coleman 
 mechanism is operative. 
 Further, we argued in  \cite{aw1,aw2}  that the space topology should be toroidal,
 and thus that the spatial curvature term $\Om_{\rm K} =0$. A new coupling constant, denoted $B$, is present in 
 our modified Friedmann equation. 
 It is proportional to 
 the coupling constant which in our underlying string field theory model describes the creation of baby universes and wormholes.
 At late times the $B$-term acts effectively as a cosmological term and one can show that 
 \beq\label{2.3}
 a(t) \;\propto\;  \e^{(27 B/4)^{1/3} t} \quad {\rm for} \quad t \to \infty.
 \eeq
Assuming a simple matter dust system, such that the pressure $p =0$, we can write
\beq\label{2.4}
\kappa \rho = \frac{C}{a^3},
\eeq
where $C$ is a positive constant. 

We can now solve \rf{2.2} numerically assuming that $a(t) \equ 0$ at $t \equ 0$, which we define as 
the time of the Big Bang. This will determine $a(t)$ up to a constant of proportionality, and the solution will 
be parametrized by $B$. One has an expansion of $a(t)$ in powers of $B t^3$:
\bea
a_B(t) &=& \alpha\, t^{2/3}  \left(1+ \frac{9}{8} \, B t^3 + O\Big( (B t^3)^2\Big) \right)\label{2.5}\\
\dot{a}_B(t) &=& \alpha \, t^{-1/3} \left(\frac{2}{3} + \frac{33}{8} \, B t^3 + O\Big( (B t^3)^2\Big) \right)\label{2.6} \\
\ddot{a}_B (t)&=& \alpha \, t^{-4/3} \left(-\frac{2}{9} + 11 \, B t^3 + O\Big( (B t^3)^2\Big) \right)\label{2.7}
\eea
In particular we find that the Hubble parameter has the expansion
\beq\label{2.8}
H_B(t) := \frac{\dot{a}_B}{a_B} = \frac{2}{3 t}  \left( 1 + \frac{81}{16} Bt^3 + O\Big( (Bt^3)^2\Big) \right)
\eeq
and that the time $t_*$ where the acceleration of the universe starts, defined as the 
zero of $\ddot{a}(t)/a(t)$, is determined by 
\beq\label{2.9}
\frac{\ddot{a}_B(t_*)}{a_B(t_*)} = -\frac{2}{9 t_*^2} \left ( 1 - \frac{405}{8} \, B t^3_* + O\Big( (Bt^3_*)^2\Big) \right) = 0
\eeq
Of course we do not expect the matter dust approximation \rf{2.4} to be  good for small $t$. At some point it has 
to be replaced by the radiation dominated equation of state and for even smaller $t$ some kind of inflationary scenario 
should presumably take place. In particular, when the universe  approaches the  Planck size 
one will have to address the underlying quantum theory of gravity. Our model based on a certain $W_3$ algebra
 is an attempt to deal with issues at this small scale, but it also led to the  modified Friedmann equation
\rf{2.2} which should be valid only when the universe is of macroscopic size. In the following we will only 
apply it at times equal or larger than the  time $t_{\rm LS}$ of last scattering of the CMB, which is well beyond the 
time when the matter dominated era starts.

\section{From CMB  H$_{\rm \bf 0}$ to standard candle H$_{\rm \bf 0}$}

Let us compare eq.\ \rf{2.8} with the corresponding equation using the standard Friedmann equation \rf{2.1}.
In that case we obtain 
\beq\label{3.1} 
H_\Lam(t) =  \frac{2}{3 t} \, \left( 1 + \oq \Lam t^2 + O\Big( (\Lam t^2)^2\Big) \right)
\eeq
As indicated by the expansions \rf{2.8} and \rf{3.1}  the fall-off of $H_B(t)$ with time $t$ is slower than that of $H_\Lam(t)$. 
The measurement of $H(t)$ using CMB refers to $H_\Lam(t)$ at the time of last scattering $t_{\rm LS}^{\rm CMB}$, 
extrapolated to the present time $t_0^{\rm CMB}$ using the $\Lam$CDM model. 
This leads to $H_{\Lam^{\rm CMB}}(t_0^{\rm CMB}):= H^{\rm CMB}_0$ \cite{planck}:
\beq\label{3.2}
67.27 - 0.60  < H^{\rm CMB}_0 < 67.27 + 0.60,\qquad t_0^{\rm CMB} = 13.8\times 10^9 \, {\rm y}.
\eeq
On the other hand the data from standard candles (SC)  such as Type Ia supernovae  and gamma-ray bursts \cite{SNIa} 
are measuring $H(t)$ with $t$ approximately equal to the present time  and the result is 
\beq\label{3.3}
73.20 - 1.30 < H^{\rm SC}_0 < 73.20 + 1.30.
\eeq
The redshift is defined as 
\beq\label{3.6}
z(t) := \frac{a(t_0)}{a(t)} -1, 
\eeq
where $t_0$ denotes the present time. Two of the simplest observables we have at our disposal are the {\it measured} 
temperature of the CMB, $T^{\rm CMB}$, and the {\it measured} Hubble constant, i.e. $H^{\rm SC}_0$. 
We assume that the measured temperature $T^{\rm CMB}=T(t_0)$ at the present time $t_0$ originates from 
 the temperature at last scattering, $T(t_{\rm LS})$ cooled down by the expansion of the universe. 
The time of last scattering is so early in the history of the universe that we will ignore the dependence of $t_{\rm LS}$ on 
parameters $\Lam$ in the $\Lam$CDM model and $B$ in our model, i.e.\ we assume 
$t_{\rm LS}^{\rm CMB}= t_{\rm LS}^{\rm B} ~( = t_{\rm LS})$. 
Finally, the fact that CMB spectrum is a black-body spectrum implies that 
the scale factor $a(t) \propto 1/T(t)$ for $t$ larger than $t_{\rm LS}$. We know $T(t_{\rm LS})$ to good precision for  physical reasons
not related to any specific cosmological model. Thus a determination of the CMB temperature $T^{\rm CMB}$
is also a determination of 
$a(t_0)/a(t_{\rm LS})$ and therefore of the redshift $z(t_{\rm LS})$. The measurement of the CMB temperature leads to 
$z(t_{\rm LS}) = 1090$. We have redshifts $z_B(t)$ and $z_\Lam(t)$ which depend  on  the parameters $B$ and 
$\Lam$ for our model and the $\Lam$CDM model, respectively. We now determine the parameters $B$ and the present 
time $t_0^{\rm B}$ for our model by requiring that  
 \beq\label{3.7}
z_{\rm B}(t_{\rm LS}) = 1090 \quad \big( = z_\Lam (t_{\rm LS})\big)
\eeq
and 
\beq\label{3.8}
H_B(t_0^{\rm B}) = H^{\rm SC}_0.
\eeq
We obtain
\beq\label{3.9}
 \big(t_0^{\rm B}\big)^3 B =0.1500 \pm 0.0056, \qquad t_0^{\rm B} = 13.882  \pm 0.043 \;[ {\rm Gyr} ].
\eeq
  
\begin{figure}[t]
\vspace{-1cm}
\centerline{\scalebox{1.0}{\rotatebox{0}{\includegraphics{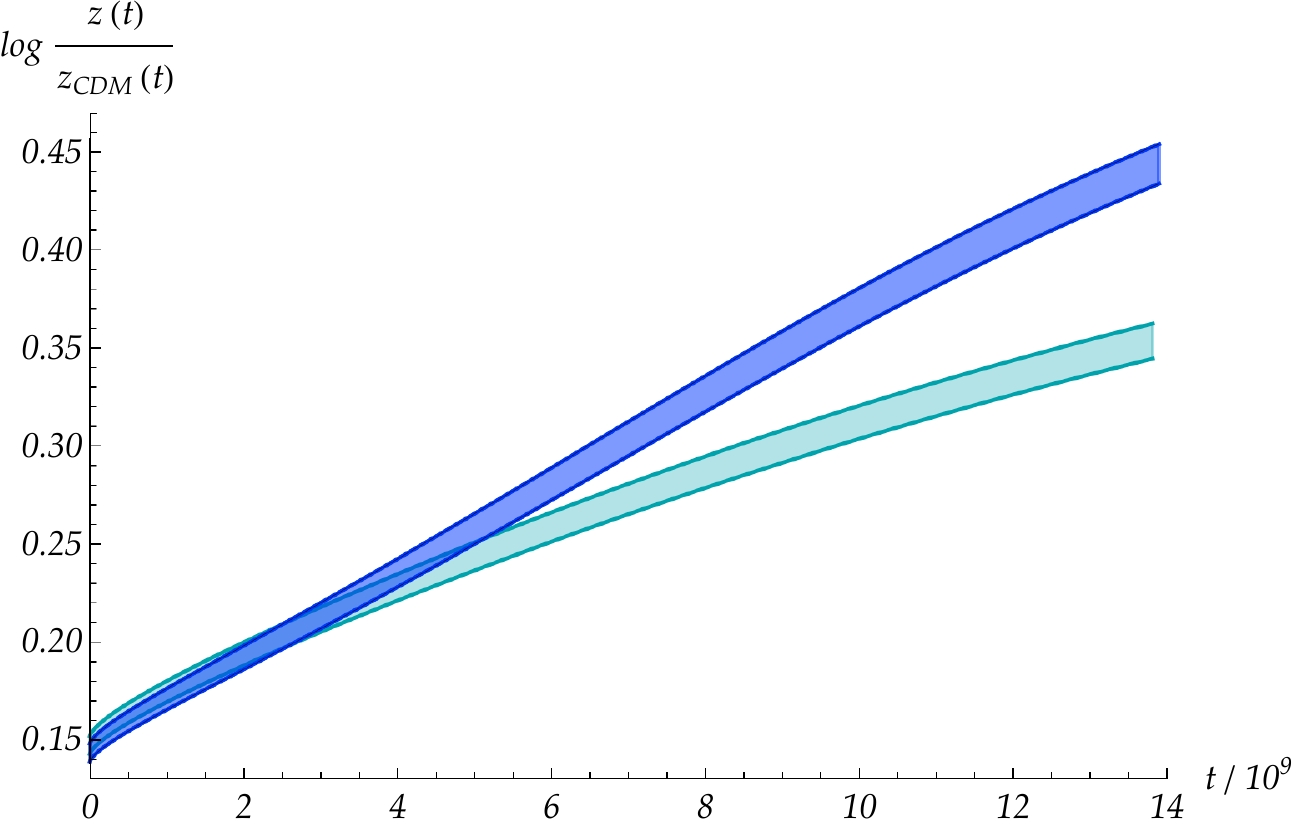}}}}
\caption[fig1]{{\small
The green  region is  $\log z_\Lam(t)/z_{\rm CDM}(t)$ of the $\Lambda$CDM model,   
with the values   of $H_0^{\rm CMB}$, $\Lam^{\rm CMB}$ and $t_0^{\rm CMB}$ taken from \cite{planck}. 
The blue lines define  the region for
$\log z_B(t)/z_{\rm CDM}(t)$ of our model, where $B$ iand $t_0^{\rm B}$ are determined  as described in the main text.  
 Note that $t_{\rm LS}$ cannot be distinguished from $t=0$ on the figure. Similarly $t_0^{\rm CMB}$ cannot be distinguished 
 from $t_0^{\rm B}$. 
}}
\label{fig1}
\end{figure}

\begin{figure}[t]
\vspace{-1cm}
\centerline{\scalebox{.9}{\rotatebox{0}{\includegraphics{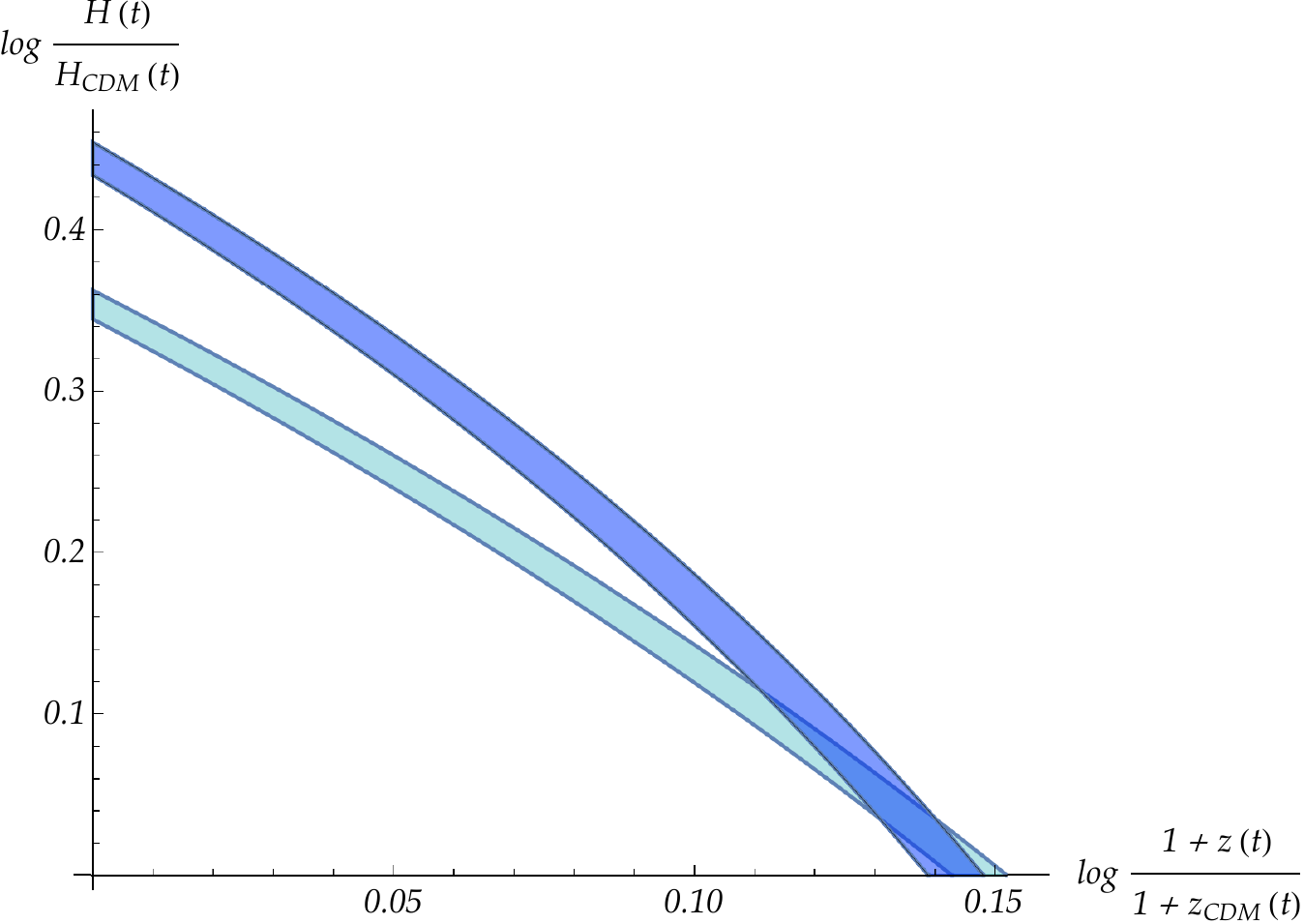}}}}
\caption[fig2]{{\small
We plot $\log \frac{H(t)}{H_{\rm CDM}(t)}$ for $H(t) = H_B(t)$ and  $H_{\Lam^{\rm CMB}}(t)$ 
against $\log \frac{1+z(t)}{1+z_{\rm CDM}(t)}$, where $z(t) = z_B(t)$ and
$z_{\Lam^{\rm CMB}}(t)$, respectively. The color code is the same as in Fig.\ \ref{fig1}.
 By construction we have  $\log \frac{H(t)}{H_{\rm CDM}(t)} \approx 0$
at $t_{\rm LS}$.
}}
\label{fig2}
\end{figure}

We have illustrated the situation in Fig.\ \ref{fig1}. Defining $z_{\rm CDM}(t):= z_{\Lam =0}(t)=z_{{\rm B}=0}(t)$, a convenient 
variable is $\log \big[z(t)/z_{\rm CDM}(t)\big]$ and we can now plot this variable for the $\Lam$CDM model 
for the CMB values of $H^{\rm CMB}_0$ given by \rf{3.2}, 
as well as for modified Friedmann values, where $B$ is in the range defined in \rf{3.9}. 
The estimate of the uncertainty of $B$ and $t_0$ 
in \rf{3.9} comes from the requirement that  $z_B(t_{\rm LS})$ should be within the limits defined by 
$z_\Lam(t_{\rm LS})$\footnote{The reason that the curves of  
$\log \big[z_B(t)/z_{\rm CDM}(t)\big]$ and $\log \big[z_{\Lam^{\rm CMB}}(t)/z_{\rm CDM}(t)\big]$  do not coincide at $t_{\rm LS}$ is that $t_0^{\rm B}$ and $t_0^{\rm CMB}$ are slightly different and thus $z_{\rm CDM} (t)$ will be different for the two models.}.
If we insist  that the central value of $H_B(t_0)$  is equal to the central 
value of $H^{\rm SC}_0$ then the uncertainty of $H_B(t_0)$ obtained in this way from the CMB data is
less than the uncertainty of  $H^{\rm SC}_0$. Fig.\ \ref{fig2} shows  plots of $H_B(t)$ and $H_{\Lam^{\rm CMB}}(t)$ against 
$z_B(t)$ and $z_{\Lam^{\rm CMB}}$, respectively.

\begin{figure}
\vspace{-1.3cm}
\centerline{
\scalebox{0.9}{\rotatebox{0}{\includegraphics{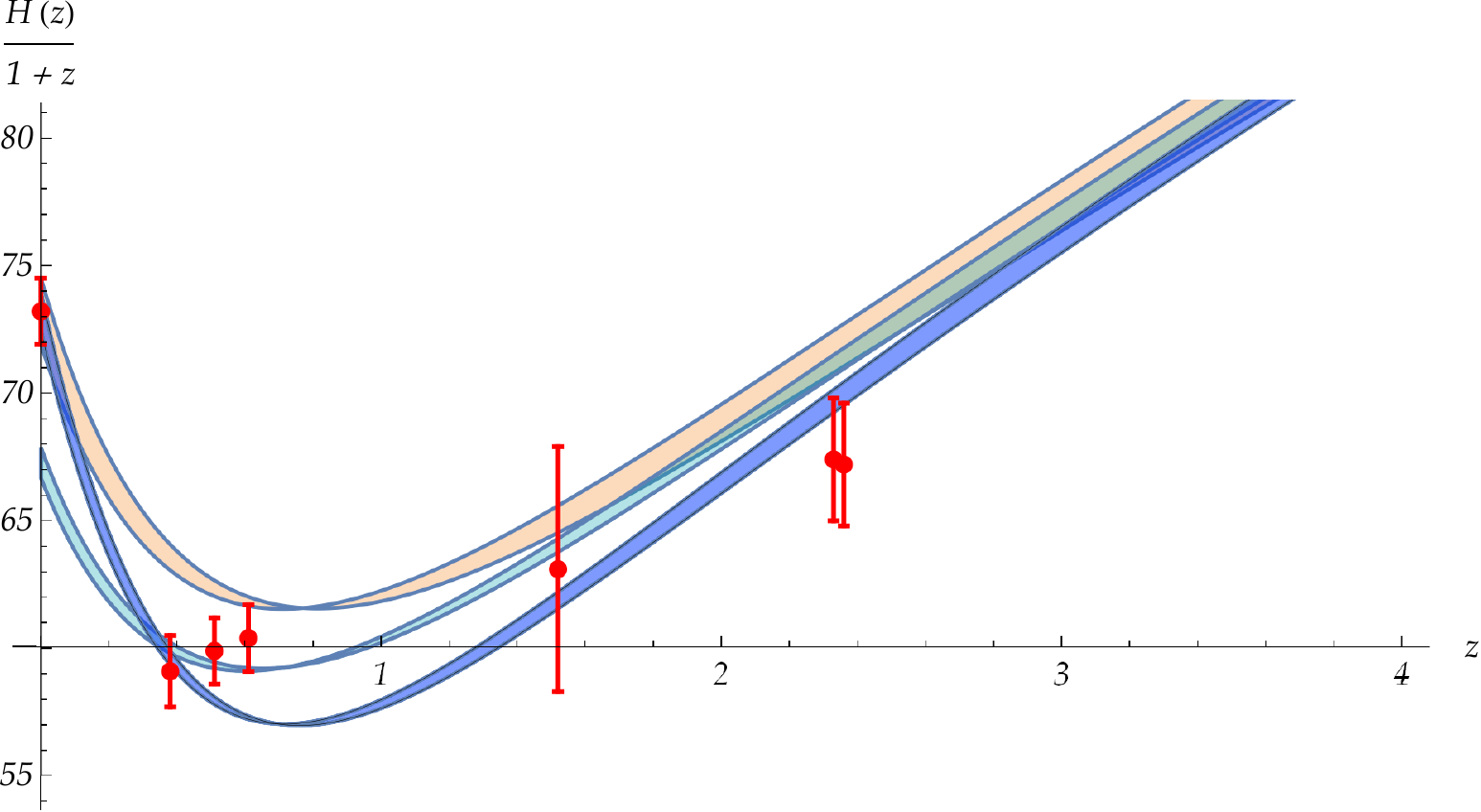}}}}
\vspace{-0.2cm}
\caption[fig3]{{\small
The blue area shows $H(z)/(1+z)$ for the modified Friedmann equation, while 
the green  area  shows $H(z)/(1+z)$  based on $\Lambda$CDM model with $H_0^{\rm CMB}$  and $\Lambda^{\rm CMB}$ taken
from \cite{planck}. The orange area shows $H(z)/(1+z)$  based on $\Lambda$CDM model with $H_0^{\rm SC}$  and 
$\Lambda^{\rm SC}$ as described in the main text.
Inserted in red are (apart from the SC data at $z \approx 0$)
also independent data from other observations: first three points from the baryon acoustic oscillation data \cite{BAO}, 
the next from quasars \cite{quasars} and the last two data points from Ly--$\alpha$ measuments \cite{lyalpha1,lyalpha2}. }}
\label{fig3}
\end{figure}

The CMB values $H_0^{\rm CMB}$ and $t_0^{\rm CMB}$ referred to in \rf{3.2} are obtained from an analysis of the complete
CMB data set \cite{planck}. This cannot be compared to our simple calculation where the only data we used from the CMB is 
the temperature $T(t_0)$ measured today. To ``compare'' the $\Lam$CDM model with our modified Friedmann
equation one can choose to determine the 
two free parameters in the $\Lam$CDM model, $\Lam$ and $t_0$, using $z_\Lam (t_{\rm LS}) = 1090$ and 
$H_\Lam(t_0) = H_0^{\rm SC}$,  in the same way as we determined $t_0^{\rm B}$ and $B$. 
Denoting the time $t_0$ determined this way $ {t}_0^{\rm SC}$,  we find
\beq\label{3.10}
{t}_0^{\rm SC} = 13.3 \times 10^9 \, {\rm y}.
\eeq
${t}_0^{\rm SC}$  is quite a lot shorter than $t_0^{\rm CMB}$. In fact one would say that ${t}_0^{\rm SC} $ is uncomfortably short
compared to the estimated oldest age of stars. 
The difference in $H(z)$ between our model based on the the modified Friedmann equation and the $\Lam$CDM model, 
both with parameters determined by fitting to $z(t_{\rm LS}) = 1090$ and $H(t_0) = H_0^{\rm SC}$, is shown 
 for small values of $z$ in Fig.\ \ref{fig3} (blue and orange regions, respectively).  
 For comparison we have also shown $H(z)$ as determined from the best fit to 
 the CMB data (green region). All curves approximately agree for large $z$, as is also seem on the figure.
 We have also included other independent measurements of $H(z)$.  
The error bars\footnote{For a given value of $z > 0$, the width between the upper and lower curves, bounding a given 
color area, does not reflect an error bar, contrary to the error bars indicated for the independent data.
The two curves correspond to the maximum and minimum of $H_0^{\rm SC}$ given by \rf{3.3} 
(for the blue and orange areas) and of $H_0^{\rm CMB}$ given by \rf{3.2} (for the green area).}  of the independent measured $H(z)$ are  quite large, and the data points few, but if we make the simplest $\chi^2$ fit to test how well the theoretical curves agree with 
the data we obtain\footnote{As usual the reduced $\chi^2_{\rm red} = \chi^2/K$, where $K$ is the number of degrees of freedom, i.e.
in the present context  6 for the blue and the orange curves and 7 for the green curve.}
\beq\label{fit}
\chi^2_{\rm red}({\rm B}) = 1.8, \quad  \chi^2_{\rm red}(\Lam^{\rm SC}) = 3.7, \quad \chi^2_{\rm red}(\Lam^{\rm CMB}) = 4.0,
\eeq
where the three reduced $\chi^2$ values refer to the blue, the orange and the green curves in Fig.\ \ref{fig3}. The orange curve 
seems to be disfavored  compared to the blue curve. 
The main contribution to $\chi^2_{\rm red}(\Lambda^{\rm CDM})$ comes from 
a single data point, namely the point at $z \equ 0$, 
which is of course precisely the source of the $H_0$ tension \footnote{The small $z$ measurements have been analyzed
in much more detail than here in attempts to determine possible deviations from the standard $\Lam$CDM model, 
e.g. \ in \cite{lowz}}.

\section{Discussion}


 Let us note that with $B$ fixed as described
above our model gives a prediction for some of the  standard variables like $\Om_{\rm m}(t_0)$ and $w(t_0)$.
The modified Friedmann equation \rf{2.2} can be rewritten as 
\beq\label{4.1}
\Om_{\rm m}(t) + \Om_{ B}(t) = 1,\quad \Om_{\rm m}(t) := \frac{\kp \rho(t)}{3} \left(\frac{a}{\dot{a}}\right)^2, \quad
\Om_B(t) =   (1\mi F(x)) (1\plu 3 F(x)),
\eeq
where $x = B a^3/\dot{a}^3$, as in eq.\ \rf{2.2}.
$\Om_B(t)$ is replacing the usual term $\Om_\Lam(t)$ in the $\Lam$DCM model and there is no $\Om_{\rm K}$--term from 
a spatial curvature as already noted. We now define $w_B(t)$ as 
\beq\label{4.2}
w_B(t) := \frac{2q(t)-1}{3\Om_B(t)}, \qquad q(t) := -  \left(\frac{a}{\dot{a}}\right)^2\frac{\ddot{a}}{a},
\eeq
and we find, using the definition $H(t) = 100 \,h(t)$ km s$^{-1}$Mpc$^{-1}$, 
\beq\label{4.3}
\Om_{\rm m}(t_0^{\rm B}) = 0.268 \pm 0.009, \qquad q(t_0^{\rm B}) = -0.82 \pm 0.01, 
\eeq
\beq\label{4.3a}
 t_*  = (7.26 \pm 0.09) \times 10^9\, {\rm y},
\qquad  w_B(t_0^{\rm B}) = -1.205 \pm 0.006.
\eeq
The value of $\Om_{\rm m}(t_0^{\rm B})$ coming from our model is somewhat lower than 
 the  value extracted from the CMB data ($\Om_{\rm m} (t_0^{\rm CMB}) = 0.315 \pm 0.007$). 
However, our value of $\Om_{\rm m}(t_0^{\rm B}) \big(h_B(t_0^{\rm B})\big)^2$ is (almost) identical to  the corresponding CMB--value 
since  $\Om_{\rm m}(t) h^2(t) \propto \rho_{\rm m}(t)$, where $\rho_{\rm m}(t)$  denotes the matter energy density, and where we
in  the matter dominated era have   $\rho_{\rm m}(t) \propto 1/a^3(t)$. Thus, to the extent that
$\Om_{\rm m}(t_{\rm LS})$ and  $h(t_{\rm LS})$ are independent of cosmological constant parameters like $\Lam$ and $B$,   
one obtains an almost model independent 
value of  $\Om_{\rm m}(t_0) \big(h(t_0)\big)^2$ as long as one assumes the same value of $z(t_{\rm LS})$ (i.e.\ of 
$a(t_0)/a(t_{\rm LS})$).

Maybe the most important prediction is that $w_B(t_0^{\rm B}) \approx - 1.2$. This is actually in agreement with certain fits to data
(see for instance \cite{ps}, section II.3 for a recent review), but is difficult to explain from physical healthy theories 
based on small modifications of the $\Lam$CDM model, and usually  $w(t) < - 1$ is associated with the 
presence of so-called ghost fields. However,  such ghost fields are not needed in our  model 
and  $w_B(t) \to -1$ for $t \to \infty$. No Big Rip will occur for the universe despite $w_B(t) < -1$, 
and the universe will at late times expand exponentially as indicated in  eq.\  \rf{2.3}.

Our modified Friedmann equation seems to be able to produce values of $H(z)$ which are in good agreement 
with the observed values, by fitting only to $T(t_0)$ and $H_0^{\rm SC}$. 
Of course, for the model to genuine do better than the $\Lam$CDM model one should repeat what is actually 
done to test the $\Lam$CDM model, namely use the set of data from the  CMB to fit a range of cosmological parameters and 
as a side result come up with a prediction for $H_B(t_0^{\rm B})$, based only  the CMB data. If this could be done 
and the values of $t_0^{\rm B}$ and $B$ still agree with the values determined here, and the analysis of the fluctuations 
works out  as well as it does for the $\Lam$CDM model, one could say that our model has resolved the present $H_0$ tension. 
This requires a detailed understanding of how to derive in our model  the fluctuations of geometry around the assumed Friedmann geometry. Presently we only know how to do that  in the case where spacetime is two-dimensional.
The extension to four-dimensional spacetime is still work in progress.  Finally, it is also our hope that 
we will actually be able to {\it calculate} the value of the constant $B$ which appears in our modified Friedmann equation 
from the underlying microscopic model based on a certain $W_3$ algebra.

\end{document}